\documentclass[prc,showpacs,nofootinbib]{revtex4} 
\usepackage{rotating}
\usepackage{epsfig}
\usepackage{dcolumn}
\usepackage{graphicx}
\begin{document}

\title{Gamma-ray Fluxes in Oklo Natural Reactors}

\author{C. R.  Gould}
\affiliation{Physics Department, North Carolina State University, 
Raleigh, NC 27695-8202, USA}
\affiliation{Triangle Universities Nuclear Laboratory, Durham, 
NC 27708-0308, USA}
\email{chris_gould@ncsu.edu} 
\author{E. I.  Sharapov}
\affiliation{Joint Institute for Nuclear Research, 141980 Dubna, Moscow Region,
Russia}
\author{A. A.  Sonzogni}
\affiliation{National Nuclear Data Center, Brookhaven National Laboratory, Upton, NY 11973-5000,
USA}

\vspace{2cm}

\date{Nov 21, 2012}

\begin{abstract}
\begin{description}

\item {Background:} Uncertainty 
in the operating temperatures  of Oklo reactor zones impacts the precision of  bounds derived for
time variation of the fine structure constant $\alpha$.  Improved 
$^{176}$Lu/$^{175}$Lu thermometry has been discussed but its usefulness may be complicated by  
photo excitation of the isomeric state $^{176m}$Lu 
by $^{176}$Lu($\gamma,\gamma^\prime $) fluorescence. 
\item {Purpose:} We calculate prompt, delayed and equilibrium $\gamma$-ray fluxes due to fission 
of  $^{235}$U  in pulsed mode operation of Oklo zone RZ10.
\item {Methods:} We use Monte Carlo modeling to calculate the prompt flux. We use improved data libraries to estimate delayed and equilibrium spectra and fluxes. 
\item {Results:} We find $\gamma$-ray fluxes as a function of energy and derive values for the coefficients $\lambda_{\gamma,\gamma^\prime}$ 
that describe burn-up of  $^{176}$Lu through  
the isomeric  $^{176m}$Lu state. 
\item {Conclusion:} The contribution of the ($\gamma,\gamma^\prime $) channel to the
 $^{176}$Lu/$^{175}$Lu  isotopic 
ratio is negligible in comparison to the neutron burn-up channels. Lutetium thermometry is fully applicable to analyses of Oklo reactor data.
\end {description}

\end{abstract}

\vspace{1pc}
 
\maketitle

\noindent {\em\bf  1. Introduction}.

\noindent  Studies of the $^{235}$U fission product isotopic ratios from Oklo \cite{Nau91} have been undertaken by many groups investigating whether the fine structure constant $\alpha$ has changed over the 2 GY period since the reactors operated. As first pointed out by Shlyachter \cite{Shl76}, the samarium isotopic ratios are sensitive to the value of $\alpha$ through the overlap of the $^{149}$Sm  E$_0$=97.3-meV neutron resonance with the thermal and epithermal portions 
of the neutron flux in the reactor. 
While the 
majority of Oklo analyses \cite{Fuj00,Dam96,Chris06,Petr06,Oneg10},
have been consistent with no shift in the resonance energy, and
therefore no change in $\alpha$, a change has been 
argued for from  astronomical observations \cite{Webb11}.

All Oklo analyses make assumptions about the operating temperatures of the reactors. But there is as yet no agreement on what these temperatures actually were. 
Utilizing the $^{176}$Lu/$^{175}$Lu isotope ratio method to determine temperatures was recently revisited by Gould and Sharapov   \cite{Gould012} . 
The method is based on the temperature dependence of the large thermal neutron capture 
cross section of $^{176}$Lu (natural abundance 2.599 \% \cite{NNDC}), and on knowing with certainty the (small) ground state branching ratio 
for thermal neutron capture on the more abundant lutetium isotope, $^{175}$Lu (natural abundance 97.401 \%). The dominant capture branch $\sigma_{175}^{m}$  leads to a short lived isomeric state in
$^{176m}$Lu, while only
a minor branch $\sigma_{175}^{g}$ leads to the  
ground state of $^{176}$Lu.
The data from Oklo show clearly that $^{176}$Lu is 
depleted in the reactor zones. But as concluded in ref. \cite{Gould012},  the degree of depletion
will be  a reliable   indicator of the temperature only if an
improved measurement of 
$B^g(175)=\sigma ^{g}/(\sigma^{g} + \sigma^{m})$ 
is performed, and if alternate 
explanations \cite{Gould012} for depletion are ruled out.

One alternate explanation for $^{176}$Lu depletion lies in the  possibility of processing of the lutetium 
isotopes in Oklo due to photo excitation of the isomeric state in $^{176m}$Lu  by $^{176}$Lu($\gamma,\gamma^\prime$) fluorescence.  
Such a process is well known in astrophysics and is 
an important channel for burning $^{176}$Lu in stellar environments \cite{Thra010,Mohr09}. The isomeric state decays to $^{176}$Hf with a half-life of 3.6 hr, and therefore provides an alternate path for removing $^{176}$Lu.  
Here we explore whether this could have been be an effect in the $\gamma$-ray 
fluxes in the reactors, taking advantage of newly developed data libraries for fission decay chains.

In $^{235}$U thermal neutron fission, about 6.6 MeV
is released in the form of prompt $\gamma$ rays, about 6.5 MeV as $\beta$ rays, and about 6.3 MeV as $\gamma$
rays following $\beta$ emission. The sum of the delayed $\beta$ and $\gamma$ energy released during the decay of 
fission products is called
decay heat and varies
as a function of time, $f(t)$,  after a single fission event at t=0. 

Beginning with the work of Way and Wigner 
\cite{Way48}, many calculations and measurements of $f(t)$
have been performed, see Tobias \cite{Tobi80} 
and Dickens \cite{Dick86} for reviews. Typically,  $f(t)\sim
t^{-1.2}$ is found for times greater than several seconds {\footnote {Note this is not the same as the time dependence of decay heat following shutdown of a long-running reactor \cite{Glass81}.}}.
With the development of more comprehensive nuclear data libraries based on level schemes derived from high resolution Ge-detector data, summation method calculations became widely accepted. These calculations gave good agreement with measurements except at the shorter times associated high Q-value $\beta$ decays. High Q-values feed levels at high excitation that can decay by emission of weak and (or) high energy $\gamma$ rays easily missed in Ge-detector measurements. As a result, the libraries were incomplete.  Recently, Total Absorption Gamma Spectrometer (TAGS) data have been included in libraries. This has eliminated the discrepancies. In particular, Algora et al. \cite{algora10} were able to report $^{239}$Pu decay heat calculations in excellent agreement with experiments for shorter times.

We follow this approach for calculating $^{235}$U decay heat $\gamma$-spectra \cite{Sonz011}, converting to fluxes  using standard energy deposition conversion coefficients. We apply Monte Carlo modeling to calculate the prompt $\gamma$-ray flux in the Oklo reactor zone. 
With these fluxes in hand, and a model of how the reactor operated, we can then estimate photo-excitation constants $\lambda_{\gamma,\gamma^\prime}$
for burning  $^{176}$Lu through the isomeric state $^{176m}$Lu.\\

\noindent {\em\bf  2.   Prompt fission $\gamma$-ray  flux in Oklo reactor zone RZ10}.

\noindent The MCNP code \cite{MCNP4} allows  
modeling of neutron transport and also provides the energy dependence of the prompt $\gamma$-ray flux. 
We use the same  input  for the Oklo zone RZ10 
as  in our previous work \cite{Chris06}. 
The model of a reactor zone is a flat cylinder of 70 cm height, 6 m diameter, surrounded by 
a 1 m thick reflector  consisting of water saturated sandstone.
As for any reactor, Oklo criticality is determined by the  geometry and  
the  composition of the active zone.
Oklo reactor zones include uraninite ${\mathrm{UO}_2}$, gangue (oxides of different 
metals with water of crystallization) and  water.  The total density of the active core material at ancient times 
was about 3 g cm$^{-3}$ for RZ10 with only 30 wt. \% 
of UO$_2$ in the RZ10 dry ore. The hydrogen to uranium atomic ratio in our model  was 
$\frac{N_H}{N_U} = 13.0$ and the multiplication coefficient of the fresh core was k$_{eff}$=1.036. Detailed composition and neutronic parameters of the RZ10 reactor zone are given in \cite{Chris06}.
The rate of fission can be deduced from the Hidaka and Holliger model  
\cite{Hida98}, which found an average RZ10 neutron fluence of 0.65 kb$^{-1}$ over a time duration of 160 kyr.

From analysis of xenon isotope abundances in Oklo grains of aluminum
phosphate,  Meshik et al. \cite{Mesh04} concluded the reactors
operated cyclically, with reactor-on periods of about 0.5 hr (1800 s),
separated by dormant reactor-off periods of 2.5 hr (9000 s). We use
this periodicity in calculating absolute fluxes. Adapting to the pulse
mode, we find  $ 1.03 \cdot 10^{18}$ fissions for the 1800 s the
reactor is on, with a thermal power of about 18 kW. We include only $^{235}$U fission here since $^{238}$U and $^{239}$Pu fission was found by Hidaka and Holliger to contribute less than ten percent to the neutron yields.

The prompt $\gamma$-ray flux during the reactor-on period is shown as the upper line in Fig. 1. The other flux shown refers to the delayed heat and is 
discussed in the next sections.  The total prompt $\gamma$-ray flux is about $3 \cdot 10^9$ $\gamma$ cm$^{-2}$ s$^{-1}$. The electromagnetic energy per fission corresponds to about 14.7 MeV, greater than the 6.6 MeV of prompt $\gamma$-ray fission energy due to neutron capture on the materials of the reactor.

\noindent {\em\bf  3. Decay heat $\gamma$-ray spectra from fission}

\noindent 

A large number of radioactive nuclides are produced from the fission of an actinide target such as $^{235}$U.
Of relevance in describing the time evolution of the decay network are the longer-lived levels, 
ground states and isomers, here called {\em materials}.
While the reactor is operating, the {\em materials} population satisfies the following set of linearly-coupled
differential equations:

\begin{equation}
\frac{dN_i(t)}{dt}=-\lambda_i N_i(t) + \sum_k \lambda_k P_{ki} N_k(t) + r FY_i ,
\label{diffeq}
\end{equation}
where the decay constant of the material $i$ is $\lambda_i=ln(2)/t_{1/2 i}$, 
$t_{1/2 i}$ is the half-life for the material $i$, 
$P_{ki}$ is the probability that the material $k$ will populate in its decay the material $i$, 
$r$ is the fission rate and $FY_i$ is the independent fission yield for the material $i$.  
When the reactor is not operating, the equations for $N_i(T)$ are the same but with $r$=0.

For each material, we have obtained the Ge-detector $\gamma$-ray spectrum $I_i(E_{\gamma})$ from the ENDF/B-VII.1 decay data sub-library \cite{endfb71},
using a detector resolution of 2 keV (FWHM) for discrete lines and a  $\Delta E_{\gamma}$=0.5 keV binning.  The spectrum is defined 
so that $I_i(E_{\gamma}) \Delta E_{\gamma}$ gives the absolute probability of observing $\gamma$-rays with energies in the 
$(E_{\gamma} - 0.5 \Delta E_{\gamma}$ , $E_{\gamma} + 0.5 \Delta E_{\gamma})$ interval per decay of the material $i$.

During the period $t$, $t+\delta t$, the $\gamma$-ray spectrum obtained by adding the contribution of all 
the radioactive nuclides in the core is given by: 

\begin{equation}
I(E_{\gamma}, t, t + \delta t) = \sum_i \lambda_i N_i(t) I_i(E_{\gamma}) \delta t ,
\end{equation}
which can be integrated over time to obtain:

\begin{equation}
I(E_{\gamma}, t_0, t_1) = \sum_i I_i(E_{\gamma}) \int_{t_0}^{t_1} \lambda_i N_i(t)  dt .
\end{equation}

The mean electromagnetic energy (EEM), and average $\gamma$ energy of the spectrum are calculated as:

\begin{equation}
EEM(t_0,t_1)= \int I(E_{\gamma}, t_0, t_1) E_{\gamma} dE_{\gamma} ,
\end{equation}

\begin{equation}
<E_{\gamma}(t_0,t_1)> = \int I(E_{\gamma}, t_0, t_1) E_{\gamma} dE_{\gamma} / \int I(E_{\gamma}, t_0, t_1) dE_{\gamma} .
\end{equation}

Equations (\ref{diffeq}) were solved numerically using the fission yields from the JEFF-3.1 library  \cite{jeff} and the decay data from the ENDF/B-VII.1
library \cite{endfb71}.

To confirm the correctness of our procedure, we first calculated delayed $\gamma$-ray spectra
for time intervals $t$=1-1800 s and $t$=1-9000 s following a single fission event at $t$=0.
These time intervals match the cycling times in our reactor model. 
 Our $EEM$ values are shown in the second column of Table 1.
In columns three through five,  we compare to values obtained by numerical integration of $f(t)$ as given in Refs \cite{Dick86,jeff,endfb71}.  The 
last column is an average of these three reference values.  Our  $EEM$ values are about 0.40 MeV lower because they are derived from 
point-wise $\gamma$-ray spectra. If we  
use the $EEM$ values in the ENDF/B-VII.1 library, which includes the
latest TAGS measurements, we  obtain  $EEM(1,1800)$=4.03 MeV and
$EEM(1,9000)$=4.77 MeV, in agreement with the  average of other results.  
Because of the pandemonium effect \cite{Hardy77}, it is a well-known fact that one obtains a lower $EEM$ value 
when using point-wise spectra, in this case about 10\%.

\begin{table}[ht]
\caption[1]{Electromagnetic energies EEM(T) (in MeV per fission) in decay heat of $^{235}$U.}
\label{tab:eem}
\begin{tabular}{||c|c|c|c|c|c|c||} \hline\hline
Time interval, s& Present work & ENDF/B-VII.1\cite{endfb71} &Dickens\cite{Dick86}
&JEFF-3.1\cite{jeff} &Average value\\ 
\hline
 1$-$1800 & 3.60 &4.12 &4.10  &4.00 &4.07\\
 1$-$9000 & 4.34 &4.84 &4.74  &4.70 &4.76\\
\hline
\hline
\end{tabular}
\end{table}

We are interested in the $\gamma$-ray flux in the pulsed cycling mode of the
reactor operation, not just for a fission event at $t=0$. We  therefore calculated 
next the $\gamma$-ray spectra during the 0.5 hr pulse (1800 s) and during the 2.5 hr (9000 s) cooling period, assuming one fission per second during 
the reactor-on pulse.
In these calculations we took $N_i$=0, that is, a fresh core.
The spectra for a 0.15 to 10 MeV energy range are shown in Figs. 2 and 3, the former on a log scale, and the latter on a linear scale to show the fine structure of the lower energy portion of the spectrum. 
The $EEM$ value for the reactor-on 0-1800 s spectrum is 3.49 MeV/fission, with $<E_{\gamma}>$  equal to 0.732 MeV. For the reactor-off 1800-10800 s spectrum, $EEM$ is equal to 1.24 MeV/fission and 
$<E_{\gamma}>$ is slightly higher, 0.813 MeV.  Adding the spectra gives $EEM(0,10800)$ = 4.73 MeV.

Using these EEM values and knowing the number of fissions per pulse, we can now calculate the prompt, delayed and equilibrium components of
the electromagnetic energy   
for the reactor-on and reactor-off time intervals.  These values are shown in columns two through
four of Table II.  The prompt entry corresponds to the energy from prompt fission 
and neutron capture $\gamma$-rays. This is zero when the reactor is off. The delayed entries are derived from  the $\gamma$-ray decay $EEM$ values given earlier. Both prompt and delayed components   
originate from one single reactor-on  pulse of power 18 kW. 
In addition to these components, there will also be
equilibrium decay heat 
associated with fission from the N previous pulses (N $>>$1). For our purposes this can be estimated over short periods of time simply by using the standard relation for the power of the after heat compared to the power of the reactor:  P$_{heat}$ = 0.066P$_{av}$  \cite{Glass81}. 
Taking the average thermal power in our model to be P$_{av}$=3 kW, and noting only half the energy in the after heat is electromagnetic, we then have
P$_{\gamma}$ = 0.033P$_{av}$, which leads to the values shown in column four. We see the equilibrium flux within the pulse will be a factor of 1.12/3.59 = 0.31 smaller than the delayed flux within the pulse. 

The last column is the energy release per unit time summed over
all electromagnetic components.  Assuming similar spectral shapes of the
components, these latter quantities will be proportional to the
total $\gamma $-ray flux during the reactor-on and reactor-off pulses. We conclude that the 
flux in time intervals between pulses (1800$-$10800 s) will be  
about 7\% of the flux during the 1800-s reactor-on pulse.

\begin{table}[ht]
\caption[1]{Electromagnetic energies (in $10^{18}$ MeV) for the pulsed model of reactor RZ10.}
\label{tab:eem}
\begin{tabular}{||c|c|c|c||c||} \hline\hline
$\Delta T$, s & prompt &delayed & equilibrium &energy rate, $10^{15}$MeV/s \\ 
\hline
0$-$1800      &15.14 &3.59 &1.12    &11.03\\
1800$-$10800  &$-$   &1.28 &5.62    &0.77\\
\hline
\hline
\end{tabular}
\end{table}

\noindent {\em\bf  4.   Conversion of the delayed $\gamma$-ray spectra to  $\gamma$-ray  fluxes}

\noindent

To estimate the  photo-excitation    parameters $\lambda_{\gamma,\gamma^\prime}$
in $^{176}$Lu it is 
necessary to know $\gamma$-ray fluxes, not simply Ge-detector spectra. 
The prompt flux is obtained directly as output of MCNP \cite{MCNP4}. 
We get the delayed flux using 
the $\gamma$-ray dose to photon fluence conversion factors 
published  by the International Commission on Radiological Protection 
(ICRP) \cite{ICRP}.  Our conversion assumes that all $\gamma$-rays produced in Oklo reactors are
absorbed by the 
active core materials, both fertile and nonfertile. 

The conversion coefficients  k(E),   
are listed in Table III,  where the absorbed dose is in units of erg g$^{-1}$ and the photon fluence 
is in units of $\gamma$ cm$^{-2}$. A polynomial fit to these data gives 
$ k(E) = 0.114 E^{3} - 1.091 E^{2} + 5.618 E - 0.189 $ with E in MeV. 

\begin{table}[ht]
\caption[1]{Fluence-to-dose conversion coefficients $k(E_{\gamma})$ for $\gamma$-rays with energy $E_{\gamma}$ \cite{ICRP}.}
\label{tab:sigmas}
\begin{tabular}{||c|c|c|c|c|c|c|c|c|c|c|c|c||} \hline\hline
$E_{\gamma}$, keV & 100 &200& 300&400&500&600&800&1000& 1500&2000&3000&4000\\ 
\hline
$k(E_{\gamma})$, 10$^{-8}$ erg g$^{-1}$/$\gamma$ cm$^{-2}$ &0.37&0.86&1.38 &1.89 &2.38 &2.84&3.69
&4.47&6.14&7.55  &9.96&12.10\\
\hline
\hline
\end{tabular}
\end{table}

For the Oklo zone RZ10, the total 
mass  is 60 tonne. To convert the reactor-on spectrum of Fig. 2 to a flux, we therefore multiply by the energy $E_{\gamma}$, divide by
$k(E_{\gamma})$, divide by the total mass of the reactor zone and the 1800 s accumulation time,  
and normalize to the $^{235}$U fission rate $ 5.71 \cdot 10^{14} s^{-1}$. 
We multiply this flux by 1.31 to take into account the contribution of the equilibrium flux.   
The resulting reactor-on delayed + equilibrium flux, binned in 100-keV intervals, is shown in Fig 1 and can now be compared to the prompt flux. 
We see it is typically an order of magnitude smaller. However,  at a few energies it does actually exceed the prompt flux. \\

\noindent {\em\bf  5.   Implications of $\gamma$-ray flux estimates for $^{176}$Lu/$^{175}$Lu thermometry}.

\noindent Lutetium thermometry is based on the
dependence of the $^{176}$Lu/$^{175}$Lu isotopic ratio on the
operating temperature of a reactor in which  $^{176}$Lu is burned (and
partially restituted) by  neutron capture reactions.
As detailed in Ref. \cite{Gould012}, this process is not described by
a single exponential decay constant. However, to set a time scale for judging the impact of the $\gamma$-ray flux, we can
introduce an effective  constant $\lambda_{n}^{eff}= 3.7\cdot 10^{-13}$
 s$^{-1}$ (effective half-life of 60 kY) based on the factor 6.4 reduction in the isotopic ratio over the 160 kY operating time of RZ10 
\cite{Chris06}. Then we write the total
 constant for disappearance of  $^{176}$Lu as  $\lambda =
 \lambda_{n}+\lambda_{\gamma,\gamma^\prime}$, where the second term
corresponds to the photo-excitation process as an alternate
explanation for $^{176}$Lu depletion. 
We omit the beta decay constant for $^{176}$Lu  $
\lambda_{\beta} = 5.8 \cdot 10^{-19}$ s$^{-1}$ because it is by six  order of magnitude 
less than  $\lambda_{n}^{eff}$.

Experimental data \cite{Thra010,Mohr09} confirm that  long lived  $^{176}$Lu in the  photon bath of celestial bodies  can be partially transformed into 
metastable $^{176m}$Lu by photons with energies around 880, 1060, 1330, and 1660 keV. Higher energy photons may also contribute \cite{Carr91}. These  photon energies correspond to excited states 
of $^{176}$Lu with specific spins and parities which allow them to act as   mediators  for 
photo excitation of the isomeric state. The rate $\lambda_{\gamma,\gamma^\prime}(E_i)$ of photo excitation of the isomeric state $^{176m}$Lu  in photon inelastic scattering 
through an intermediate state (IS) at energy  $E_i$ is  given by 

\begin{equation}
\lambda_{\gamma,\gamma^\prime}(E_i)= \int \Phi_{\gamma}(E) \sigma_{\gamma,\gamma^\prime}(E,E_i) dE = \Phi_{\gamma}(E_i) \sigma^{int}_
{\gamma,\gamma^\prime}(E_i).
\end{equation}
Here $\Phi_{\gamma} (E_i)$ is the differential photon flux at energy $E_i$ having units of keV$^{-1}$ cm$^{-2}$ s$^{-1}$, and 
$\sigma^{int}_
{\gamma,\gamma^\prime}(E_i)  =  \int \sigma_{\gamma,\gamma^\prime}(E,E_i) dE $ 
is the integrated  cross section for the IS. When several IS contribute, $\lambda_{\gamma,\gamma^\prime}$ will be a sum over the individual IS contributions.

In writing the rate on the right as the product of two factors we are implicitly assuming the flux is continuous, and varying slowly over the narrow resonance energy $E_i$. This is true in stellar environments, and for example, in bremsstrahlung experiments. It is not necessarily the case in our situation where the photon spectra are made up from a sum over many discrete $\gamma$-ray lines. However, our fluxes propagate in the dense environment of the Oklo reactors, and as discussed by von Neumann-Cosel et al. \cite{vN91}, Compton scattering can broaden otherwise discrete spectra quite significantly. The MCNP code models Compton scattering, pair-production and electron bremsstrahlung fully, leading to the continuous prompt spectrum shown in Fig. 1. The delayed spectra show more structure, but for purposes of estimating upper bounds on the photo-excitation process we assume our 100 keV averaging procedure will serve as a useful approximation.

Reported values of the photo excitation 
cross sections given in the literature include 
$\sigma^{int}_
{\gamma,\gamma^\prime}(E_\gamma)$=33.4 mb$\cdot$eV for $E_\gamma$=839 keV 
(this is an upper limit in Ref. \cite{Mohr09}, for a laboratory environment, as opposed to a fully ionized environment), and higher values  
$\sigma^{int}_
{\gamma,\gamma^\prime}(E_\gamma)$=140 mb$\cdot$keV and 350 mb$\cdot$keV, for 4- and 6-MeV bremsstrahlung irradiations respectively, 
with an assumed IS energy of 2.125 MeV \cite{Carr91}.

With these cross sections, and the  
fluxes $\Phi_{\gamma} (E)$ of Fig. 1 at hand, we can now calculate
$\lambda_{\gamma,\gamma^\prime}$ for Oklo reactor RZ10. 
Taking into account the pulse structure, we use a weighted average flux consisting of 1/6 of the sum of the prompt, delayed and equilibrium fluxes while the reactor is on, and 5/6 of 7\% of this sum while the reactor is off.  

The option of $E_\gamma$=839 keV, with a total spectral flux of
$0.86 \cdot 10^6$ $\gamma$ cm$^{-2}$ s$^{-1}$ keV$^{-1}$, 
leads to  
 $\lambda_{\gamma,\gamma^\prime} = 0.26 \cdot 10^{-22}$ s$^{-1}$ which is negligible.  
The 6-MeV bremsstrahlung option with  $E_\gamma $ = 2.1 MeV, has a total spectral flux
of $1.9 \cdot 10^5$ $\gamma$ cm$^{-2}$ s$^{-1}$ keV$^{-1}$  and gives 
$\lambda_{\gamma,\gamma^\prime} = 6.8 \cdot 10^{-20}$ s$^{-1}$.  
However, even this value is less
than the 
$\lambda_{n}^{eff}$ by seven orders of magnitude.  We conclude that destruction of
$^{176}$Lu in the Oklo reactors is not 
influenced by any part of the reactor-on photon flux. 

After complete shutdown, the power of the $\gamma$-ray decay heat   
decreases approximately as
$t^{-0.3}$ \cite{Glass81}, reaching
about 1\% of the 
equilibrium value after a year. At this point it can be ignored. We
conclude therefore that the decay heat is also not able to change the
lutetium isotopic ratio even over the long period the reactor has been shut down.

The particular values we have found are specific to the uniform pulsed
mode of operation that we have assumed. 
However, the burn rate for steady state reactor
operation will not be much different since our result is  
determined mainly by the prompt flux,  which scales
inversely with the live time of the reactor. 
Absent a many orders of magnitude larger photo-excitation cross section through
as yet undetermined levels, 
we see the photon intensities in the reactor  are insufficient to
alter the  $^{176}$Lu/$^{175}$Lu  
isotopic ratios associated with neutron transmutation.  
\\

\noindent {\em\bf  6.  Conclusions}

\noindent We have developed realistic models of the prompt and delayed
$\gamma$-ray  fluxes in Oklo 
natural nuclear reactors, taking advantage of recent releases of the
databases and methodology that 
accurately represent short time decay heat in fission processes. We
have compared $^{176}$Lu transmutation 
rates associated with photo excitation to transmutation rates
associated with neutron capture processes. 
In contrast to astrophysical processes, we find $^{176}$Lu/$^{175}$Lu
isotopic ratios in Oklo are  not 
influenced by decay heat electromagnetic radiation, either during
reactor operation, or after reactor shutdown. 
Lutetium thermometry, as recently studied \cite{Gould012},  
is therefore  applicable to analyses of Oklo reactor data.

\acknowledgments

This work was supported by the US Department of Energy, Office of Nuclear Physics, under Grant No. DE-FG02- 97ER41041 (NC State University), and under contract No. DE-AC02-98CH10886 with Brookhaven Science Associates (NNDC).

\begin{figure}
\begin{center}
\includegraphics[width=7.5in]{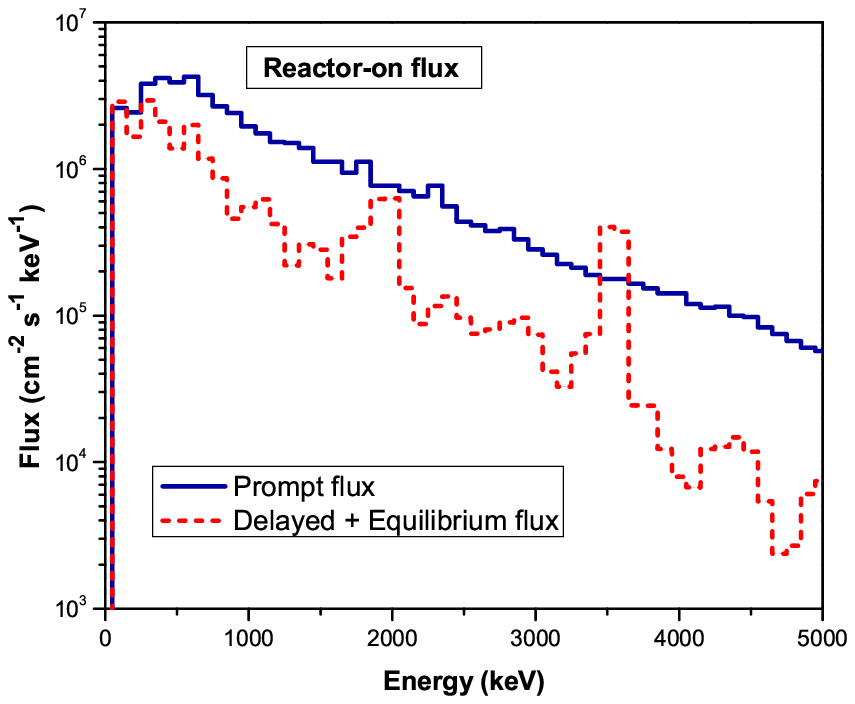}
\caption{(Color online). Prompt and delayed $\gamma$-ray fluxes 
$\Phi_{\gamma} (E)$
in Oklo reactor zone RZ10 during the period the reactor is on. The fluxes are calculated for an 18 kW reactor cycling on for 0.5 hr, and off for 2.5 hr. The prompt flux, from MCNP, is the upper line. The lower line is the delayed flux for one 0.5 hr fresh-core reactor-on pulse, multiplied by 1.31 to take into account the equilibrium flux  associated with the N previous reactor-on pulses (N $>>$1). The statistical uncertainty in the prompt spectrum simulation is 5 \%. The structure in the delayed flux is due to incomplete averaging of contributions from the discrete lines shown in Fig. 3.  }
\end{center}
\end{figure}

\begin{figure} 
\begin{center}
\includegraphics[width=7.5in]{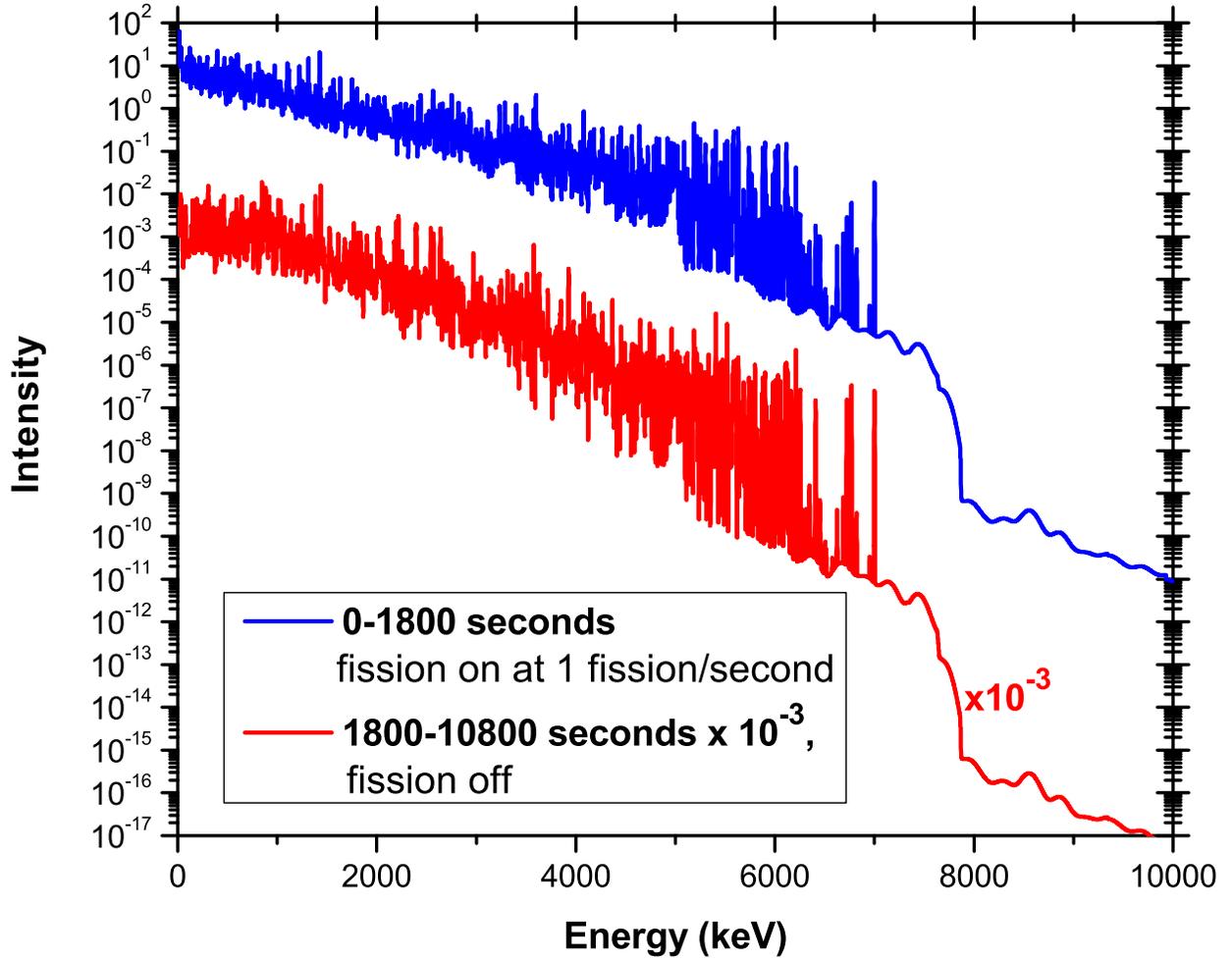}     
\caption{(Color online). Time-integrated $\gamma$-ray spectra in the 0.15 MeV to 10 MeV range for the $T=0-1800$ s 
reactor-on  pulse (upper)at one fission/sec, and for the reactor-off period $T=1800-10800$ (lower), the latter scaled down by a factor of $10^3$ for clarity.  Counts are per 1-eV energy interval.}
\end{center}
\end{figure}

\begin{figure} 
\begin{center}
\includegraphics[width=7.5in]{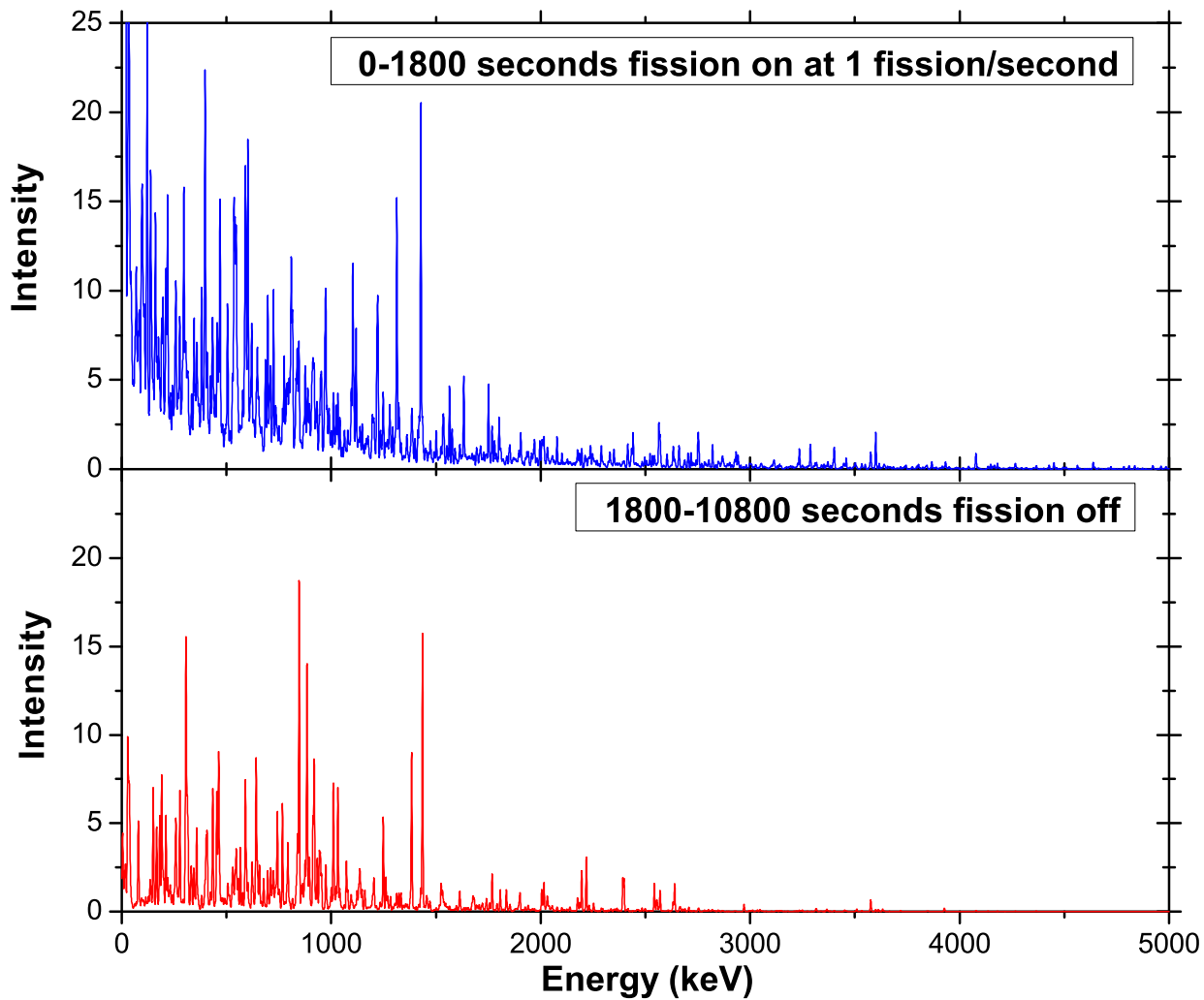}     
\caption{(Color online). Time-integrated $\gamma$-ray spectra in the 0.15 MeV to 5 MeV range for the $T=0-1800$ s 
reactor-on  pulse (upper), and for the reactor-off period $T=1800-10800$ (lower).  Counts are per 1-eV energy interval. The discrete nature of the spectra is more evident as compared to Fig. 2, but (see text) can still be treated as quasi-continuous for purposes of estimating photo-excitation probabilities.}
\end{center}
\end{figure}
\vspace{1cm}

\end{document}